\documentclass{ieeeaccess}
\usepackage{cite}
\pdfoutput=1
\usepackage{amsmath,amssymb,amsfonts}
\usepackage{algorithmic}
\usepackage{algorithm}
\usepackage{verbatim}
\usepackage{cite}
\usepackage{subfigure}
\usepackage{makecell}
\usepackage{longtable}
\usepackage{pifont}
\usepackage{graphicx}
\def\BibTeX{{\rm B\kern-.05em{\sc i\kern-.025em b}\kern-.08em
    T\kern-.1667em\lower.7ex\hbox{E}\kern-.125emX}}
\begin{document}
\history{Date of publication xxxx 00, 0000, date of current version xxxx 00, 0000.}
\doi{10.1109/ACCESS.2017.DOI}

\title{A Systematic Literature Review on Blockchain Enabled Federated Learning Framework for Internet of Vehicles}

\author{\uppercase{Mustain Billah}\authorrefmark{1}, \uppercase{Sk. Tanzir Mehedi}\authorrefmark{2}  \IEEEmembership{Member, IEEE}, \uppercase{Adnan Anwar}\authorrefmark{3} \IEEEmembership{Member, IEEE}, \uppercase{Ziaur Rahman}\authorrefmark{4} \IEEEmembership{Member, IEEE}, \uppercase{Rafiqul Islam}\authorrefmark{5} \IEEEmembership{Senior Member, IEEE}}

\address[1]{Department of Computer Science Engineering, Jashore University of Science and Technology, Jashore, Bangladesh (e-mail: mu.billah@just.edu.bd)}

\address[2]{Department of Information Technology, University of Information Technology and Sciences, Dhaka, Bangladesh (e-mail: tanzirmehedi@uits.edu.bd)}

\address[3]{Centre for Cyber Security Research and Innovation (CSRI), Deakin University, Geelong 3216, Australia (e-mail: adnan.anwar@deakin.edu.au)}

\address[4]{Faculty of Sci Eng and 
Built Env (SEBE), Deakin University, Geelong VIC 3220, Australia (e-mail: r.ziaur@deakin.edu.au)}

\address[5]{School of Computing, Mathematics and Engineering, Charles Sturt University, Australia (e-mail: mislam@csu.edu.au)}



\corresp{Corresponding author: Mustain Billah (e-mail: mu.billah@just.edu.bd).}

\begin{abstract}
While the convergence of Artificial Intelligence (AI) techniques with improved information technology systems ensured enormous benefits to the Internet of Vehicles (IoVs) systems,  it also introduced an increased amount of security and privacy threats. To ensure the security of IoVs data, privacy preservation methodologies have gained significant attention in the literature. However, these strategies also need specific adjustments and modifications to cope with the advances in IoVs design. In the interim, Federated Learning (FL) has been proven as an emerging idea to protect IoVs data privacy and security. On the other hand, Blockchain technology is showing prominent possibilities with secured, dispersed, and auditable data recording and sharing schemes. In this paper, we present a comprehensive survey on the application and implementation of Blockchain-Enabled Federated Learning frameworks for IoVs. Besides, probable issues, challenges, solutions, and future research directions for BC-Enabled FL frameworks for IoVs are also presented. This survey can further be used as the basis for developing modern BC-Enabled FL solutions to resolve different data privacy issues and scenarios of IoVs.
\end{abstract}


\titlepgskip=-15pt

\maketitle
\section{Introduction}

    \PARstart{R}{ecently} conventional vehicular ad-hoc networks are progressively advancing towards the Internet of Vehicles (IoVs). IoV is a regular network system with emphasized sensing, information communication, and computational abilities such as vehicular sensors and IoT devices, Roadside Units (RSUs), etc. IoV is the core technology that has the capability to solve the current traffic problems including various smart city applications. Road security and gridlock have been serious issues and will keep on ascending with the increase in the number of vehicles. In near future, IoV is expected to be one of the core driving force to solve the aforementioned challenges. \\
	
	In an IoV network, vehicles are equipped with modern communication and sensing technologies enabling data sharing and trading among vehicles. Vehicles exchanges communicate essential security messages intermittently just like other data, for example, crash notice, path change data, crisis cautioning, latest traffic data, dynamic route, infotainment, etc. Cellular frameworks and RSUs are set adjacent to the streets. They provide street safety, routing, and administration to the other units. The enormous volume of data captured by vehicle sensors including GPS, RADAR, etc is promoting data-driven AI models.\\
	
	Deployed smart devices in IoV are portable and distributed in nature which draws in various security issues. A digital attack may be more serious in an IoV setup contrasted with different domains as it is directly related to the driver's or passenger's physical injury. If a vehicle goes under the control of a malicious attacker on the roadway, it can prompt a terrible mishap bringing about a few passings and wounds. To construct a productive and powerful Intelligent Transport System (ITS), a learning system should be set up, which does not just give street safety and other traffic-related administrations only, additionally has the option to distinguish any sort of inconsistencies and interruption and take remedial measures. Conventional techniques for countering security issues ensure safety measurements solely after the event of explicit sorts of attacks. However,  the sorts and examples of attacks recently have changed radically. Attacks utilizing polymorphic viruses can not be easily recognized and predicted as their signatures change continuously. Thus, the conventional Machine Learning (ML) approach for identifying any sort of security vulnerabilities in IoV systems is drawing a lot of attention to researchers in recent years.\\
		
	Training an autonomous driving model with high accuracy in a low-latency is difficult as vehicles have a limited number of resources for computing. To tackle this issue, studies focus on autonomous driving frameworks based on  Multi-access Edge Computing (MEC) servers. In such cases, mobile devices gather information, and this information is transferred to a cloud server. The cloud server processes the information and creates inference models. Powerful MEC servers can assist autonomous vehicles to train more precise models in an acceptable latency. But frequent interaction among the end devices and the server will bring extreme channel pressure.\\
	
	\noindent Also, the original driving data need to be shared which will reveal a lot of private data including the identity of drivers, their standard daily practice, and behavior preference, which is actually inverse to the privacy requirements. The number of autonomous vehicles is increasing day by day. Again, all the managerial decisions must be taken within a confined time-frame. So, a centralized cloud-based methodology can't offer scalability and acceptable latency. Another additional challenge for vehicular networks is that a centralized framework requires full connectivity.\\
	
	To adapt to the rise in probable privacy and security issues, the centralized ML paradigm is moving towards a more decentralized and distributed learning framework, specifically in a Federated learning setup. To mitigate the privacy risks Google \cite{mcmahan2017communication} proposed FL, where a model is jointly trained by multiple parties. A deep neural network model is trained by the central server with the cooperation of multiple clients also called workers. The central server initially spreads an underlying training model to the clients. Each client based on this model and its local dataset calculates local updates (e.g. Stochastic Gradient Descent (SGD)) of the global model. All clients, after a predefined training period, send their own updates to the central server. The central aggregate the local models to construct a global model. However, these steps are kept iterating until an acceptable global model is achieved. Thus data privacy is ensured, as the clients do not send and store the local datasets at the central server. FL is expected to provide more advantages than the traditional decentralized learning approaches: Followings are some advantages provided by FL:
	
	\begin{itemize}
		\item FL can handle the unbalanced distribution of data. It can properly execute with Non-IID data, where, existing decentralized methodologies only accept IID data.
		
		\item In FL, clients only share the local updates of the global model with the central server. Thus, FL reduces communication overhead. Also, it can determine the selection of clients. Thus communication efficiency and system performance improvement are also ensured.
		
		\item A large number of clients should participate in achieving high accuracy in deep neural network models. With some wonderful features like privacy preservation and low communication overhead, FL can easily involve such large engagement.
	\end{itemize}
	
		\begin{table*}[!ht]
		\centering
		\caption{Comparison of ML, FL, and  BC-Enabled FL models on the basis of Central Aggregator, Economic Modeling, Adversary Control, and Privacy}
		\label{tab:comparison among MLFLBCFL}
		\begin{tabular}{|p{3cm}|p{3cm}|p{3cm}|p{3cm}|p{3cm}|}
			\hline
			 & \textbf{Central Aggregator} & \textbf{Economic Modeling} & \textbf{Adversary Control} & \textbf{Privacy} \\ \hline
			\textbf{Machine Learning} & Requires central aggregator & Lacks economic modeling & No mechanism for adversary control &  No data privacy and no model security\\ \hline
			\textbf{Federated Learning}  &   Requires central aggregator  & Lacks economic modeling &Requires adversary control  &  Only provide privacy, not model security\\ \hline
			\textbf{Blockchain Enabled Federated Learning} & No requirement of third party & Cryptocurrency based incentives & Security through smart contract and inherent security mechanisms by BC&  Provide both privacy and data/model security \\ \hline
		\end{tabular}
	\end{table*}
	
	These benefits have highlighted the importance of FL applications in different areas, including smartphones, supply chain management, Intelligent medical system, finance, etc. Vehicular IoT frameworks include a huge amount of vehicle sensor information and different sorts of utilizations in complex vehicular systems. In such cases, the Quality-of-Service (QoS) of the end-users must be supported by optimally utilizing the limited storage and computing capacity and communication resources \cite{yoshinagaspatial}. In the meantime, demand for various autonomous and intelligent transport services is rising. These services require fast reaction, high dependability, and precision. A few services face an outrageous change in their asset requests concerning time and area. Current IoV frameworks just consider the intelligence of a solitary vehicle agent. Moreover, it relies upon the storing of vehicle information in the cloud. Thus it can not fulfill the necessities of such arising services. Furthermore, vehicles are outfitted with various kinds of sensors that produce privacy-sensitive data. Also, the conditions change continuously with time and street types. FL can coordinate all these necessities, efficiently utilize vehicle computing capabilities, and secure the local data.\\
			
	Though FL gives awesome security to the learning structures, it actually works based on a centralized aggregator. Moreover, it requires an economic model to attract mobile devices in the training process. Again, a malignant vehicle may cause a poisoning attack and alter information. Again, a self-centered vehicle may not accommodate information assortment bringing about wrong weights of a local model. So, considering the probability of such potential attacks in FL, BC is being utilized with FL to give a decentralized arrangement, for controlling incentives and dependably guaranteeing security and protection. Table \ref{tab:comparison among MLFLBCFL} illustrates the comparison of ML, FL and BC enabled FL on the basis of central aggregator, economic modeling, adversary control, and privacy.

	\subsection{Contributions}
	
	This paper aims to review the development of BC-Enabled FL frameworks for critical infrastructures. The main contributions of this paper are as follows:
	\begin{enumerate}
		\item To the best of our knowledge, this is one of the earliest survey work that surveys the possibilities of the convergence of BC and FL together for IoV applications.
	
		\item Review the development of BC-Enabled FL framework from the perspective of existing FL problems and the roles of BC in addressing the FL problems.
		
		\item We not only discuss the frameworks applying BC-Enabled FL in IoVs but also explain the recent advances in FL-based IoVs and BC-based IoVs in two separate sections.
		
		\item Sort out the current challenges and future research directions of BC-Enabled FL for IoVs.
	\end{enumerate}

	\subsection{Overview of Related Survey Articles}
	
	This survey article on BC-Enabled FL frameworks for IoVs has some distinct focus compared to the existing studies. We extensively cover the area of BC-Enabled FL in IoVs. Previous survey articles have focused on FL-based IoVs, and few of them focused on BC-based IoVs. However, to the best of our knowledge, there is no prior detailed survey article that thoroughly addresses BC-Enabled FL frameworks for IoVs. Table \ref{tab:Comparison of survey papers} shows the comparison of previous survey articles on the IoVs based on FL and/or BC.
	
	\begin{table*}[!ht]
		\centering
		\caption{Summary comparison of previous survey articles on the Internet of Vehicles (IoVs). '\ding{51}' indicates that the topic is covered, '\ding{55}' indicates that the topic is not covered, and '\ding{84}' indicates that the topic is partially covered}
		\label{tab:Comparison of survey papers}
		\begin{tabular}{|c|c|c|c|c|c|c|}
			\hline
			\textbf{Ref.} & \textbf{Year} & \textbf{\makecell{Federated\\ Learning}} & \textbf{Blockchain} & \textbf{\makecell{IoVs Based on \\Federated Learning}} & \textbf{\makecell{IoVs Based on\\ Blockchain}} & \textbf{\makecell{IoV Based on \\ Blockchain-Enabled \\ Federated Learning}}\\ \hline
			\cite{yang2019federated} & 2019	& \ding{51} & \ding{55} & \ding{55} & \ding{55} & \ding{55}\\ 
			\hline
			\cite{aledhari2020federated} & 2020 & \ding{51} & \ding{51} & \ding{55} & \ding{55} & \ding{55}\\
			\hline
			\cite{lim2020federated} & 2020 & \ding{51} & \ding{55} & \ding{84} & \ding{55} & \ding{55}\\ 
			\hline
			\cite{mothukuri2021survey} & 2020 & \ding{51} & \ding{55} & \ding{55} & \ding{55} & \ding{55}\\ 
			\hline
			\cite{jiang2020federated} & 2020 & \ding{51} & \ding{55} & \ding{55} & \ding{55} & \ding{55}\\ 
			\hline
			\cite{xu2021federated} & 2020 & \ding{51} & \ding{55} & \ding{55} & \ding{55} & \ding{55}\\ 
			\hline
			\cite{mollah2020blockchain} & 2020 & \ding{55} & \ding{51} & \ding{55} & \ding{84} & \ding{55}\\ 
			\hline
			\cite{truong2021privacy} & 2021 & \ding{51} & \ding{55} & \ding{55} & \ding{55} & \ding{55}\\ 
			\hline
			\cite{pham2021fusion} & 2021 & \ding{51} & \ding{55} & \ding{55} & \ding{84} & \ding{55}\\ 
			\hline
			\cite{wang2021survey} & 2021 & \ding{55} & \ding{51} & \ding{55} & \ding{84} & \ding{55}\\ 
			\hline
			\cite{zhang2021survey} & 2021 & \ding{51} & \ding{55} & \ding{55} & \ding{55} & \ding{55}\\  
			\hline
			\cite{hou2021systematic} & 2021 & \ding{51} & \ding{51} & \ding{55} & \ding{55} & \ding{55}\\ 
			\hline
			\cite{mendiboure2020survey} & 2020 & \ding{55} & \ding{51} & \ding{55} & \ding{51} & \ding{55}\\ 
			\hline
			
			\textbf{Proposed} & \textbf{2021} & \textbf{\ding{51}} & \textbf{\ding{51}} & \textbf{\ding{51}} & \textbf{\ding{51}} & \textbf{\ding{51}}\\ 
			\hline
		\end{tabular}
	\end{table*}
	
	\subsection{Related Components and Terms}
	
	In this section, we represent some main components and terms
	(Table \ref{tab:acronyms}) related to IoVs system:
	
	\begin{itemize}
		\item \textbf{Roadside Units (RSUs):} RSUs are distributed nodes that collect locally trained ML models in their area to train global aggregated model. 
		
		\item \textbf{Vehicles:} Vehicles are dynamic and mobile edge computing devices. Vehicles are of two types: ordinary edge vehicles and representative vehicles of the group elected by RSU.
		
		\item \textbf{Certification Authority (CA):} The certification authority provides cipher suites for ensuring secure data transmission in the communication network. Registration of vehicles and RSUs on the IoV system is completed in the centralized certification authority.
	
	\end{itemize}

	\begin{table*}[!ht]
		\centering
		\caption{LIST OF ACRONYMS AND CORRESPONDING DEFINITIONS}
		\label{tab:acronyms}
		\begin{tabular}{|l|l|}
			\hline
			\textbf{Acronyms} & \textbf{Definitions}\\ \hline
			
			MEC & Mobile Edge Computing  \\ \hline
		IID & Independent and Identically-Distributed   \\ \hline
		CNN & Convolutional Neural Network   \\ \hline
		LSTM & Long Short Term Memory   \\ \hline
		TFP & Traffic Flow Prediction   \\ \hline
		PFP & Passenger Flow Prediction   \\ \hline
		IDS & Intrusion Detection System  \\ \hline
		DMV & Department of Motor Vehicles  \\ \hline
		UAV & Unmanned Aerial Vehicle  \\ \hline
		UGV & Unmanned Ground Vehicle  \\ \hline
		AQI & Air-Quality Index   \\ \hline
		ITS & Intelligent Transport System  \\ \hline
		IoV & Internet of Vehicle  \\ \hline
		BC & Blockchain  \\ \hline
		FL & Federated Learning   \\ \hline
		ML & Machine Learning  \\ \hline
		DL & Deep Learning  \\ \hline
	    AI & Artificial Intelligence  \\ \hline
		SGD & Stochastic Gradient Descent  \\ \hline
		SVM & Support Vector Machine  \\ \hline
		QoS & Quality-of-Service  \\ \hline
		DFL & Deep Federated Learning  \\ \hline
		DRL & Deep Reinforcement Learning  \\ \hline
		PoFL & Proof-of-Federated Learning  \\ \hline
		VANET &Vehicular Ad-hoc Network  \\ \hline
			
		\end{tabular}
	\end{table*}

	\subsection{Paper Organization}
	
	The rest of this paper is structured as follows. Section \ref{Background} gives background information about FL, BC, and their underlying working processes. It also contains convergence of FL and BC. In Section \ref{BFL_IOV}, frameworks, architectures, and literature on the BC-Enabled FL paradigm for IoV are identified. Section \ref{BFL_AS} highlights some Application Scenarios of BC-Enabled FL frameworks for IoVs. In Section \ref{challenges}, major open research challenges, solutions, and possible research directions are discussed. Finally, Section \ref{Conclusion} provides the concluding remarks.

\section{Background} 
\label{Background}
	
	This article deals with FL frameworks for critical infrastructures based on BC technology. Hence, background discussion is divided into three broad categories: FL, BC, and convergence of FL and BC.

	\subsection{Federated Learning}
	
	\begin{figure*}[!ht]
		\centering
		\includegraphics[scale = 0.5]{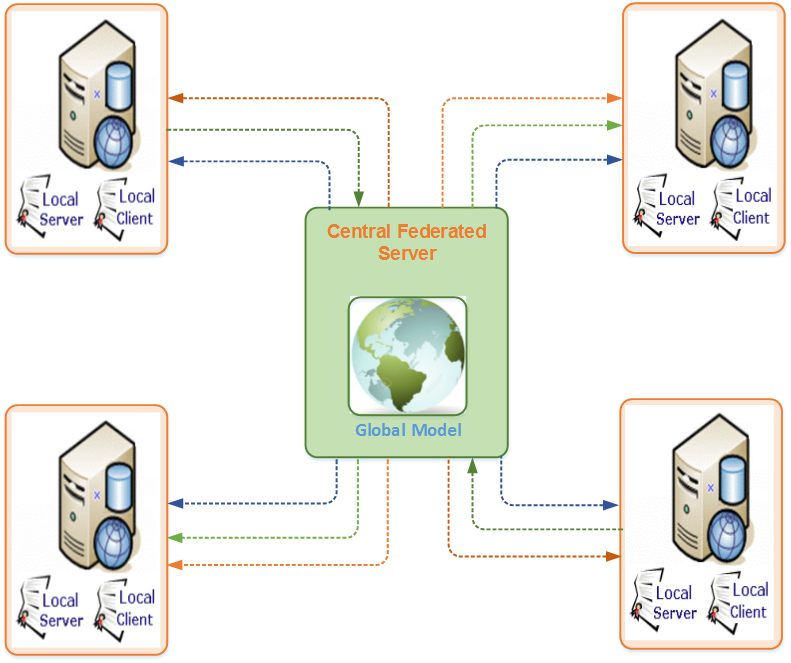}
		\label{fig:FL}
	\end{figure*}
	
	ML has been advanced by accessibility to large volumes of data, advanced computation, and DL models with an admirable achievement rate. However, the following challenges need to be addressed for wider applications and prospects: 
	
	\begin{itemize}
		\item Issues regarding client information protection, classification, and the laws that administer them.
		\item Inability to construct ML models because of insufficient data. 
		\item The computational cost of preparing an ML model.
	\end{itemize}
	
	In numerous situations, the conventional cloud-driven ML approaches are at this point not appropriate because of the difficulties of following data protection regulations and handling individual information. Given the challenges related to the high computation on such personal devices alongside the increasing privacy issues, the pattern of decentralized AI has normally risen which meets the MEC \cite{hu2015mobile} with AI/ML methods in a privacy-preserving decentralized manner.\\
	
    In such a manner, FL is a better option for the cloud-driven ML procedure that works with an ML model to be prepared cooperatively while holding unique individual information on their devices, consequently possibly mitigating privacy issues. It covers different software engineering viewpoints including ML, distributed computing, information protection, and security that empowers end-clients devices to locally prepare a common ML model on neighborhood information. Just parameters in the training cycle are traded to the aggregator. FL is an advancement to distributed learning. It works efficiently with unbalanced and non-IID data whose sizes might traverse a few significant degrees. Such heterogeneous datasets live at an enormous number of dissipating mobile devices under temperamental networks and limited communication bandwidth \cite{mcmahan2017communication}.\\
	
	FL addresses ML concerns by giving a profoundly prepared ML model without the danger of uncovering preparing information. FL additionally handles the issue of having lacking information by giving a trust factor among heterogeneous areas. Such security safeguarding procedures of FL draw the attention of various communities to use it solely, saving customer information protection and profiting advantages of having a model prepared on bigger data. FL is considered as an iterative interaction wherein every cycle, the main ML model is improved. FL executions can be summed up into the accompanying three stages: Model choice, Local model preparing, and Accumulation of local models. Figure \ref{fig:FL} pictures the FL engineering and preparing approach according to these three steps.

	\subsubsection{Federated Learning in IoVs}
	
\begin{figure*}[h!]
		\centering
		\includegraphics[scale = 0.5]{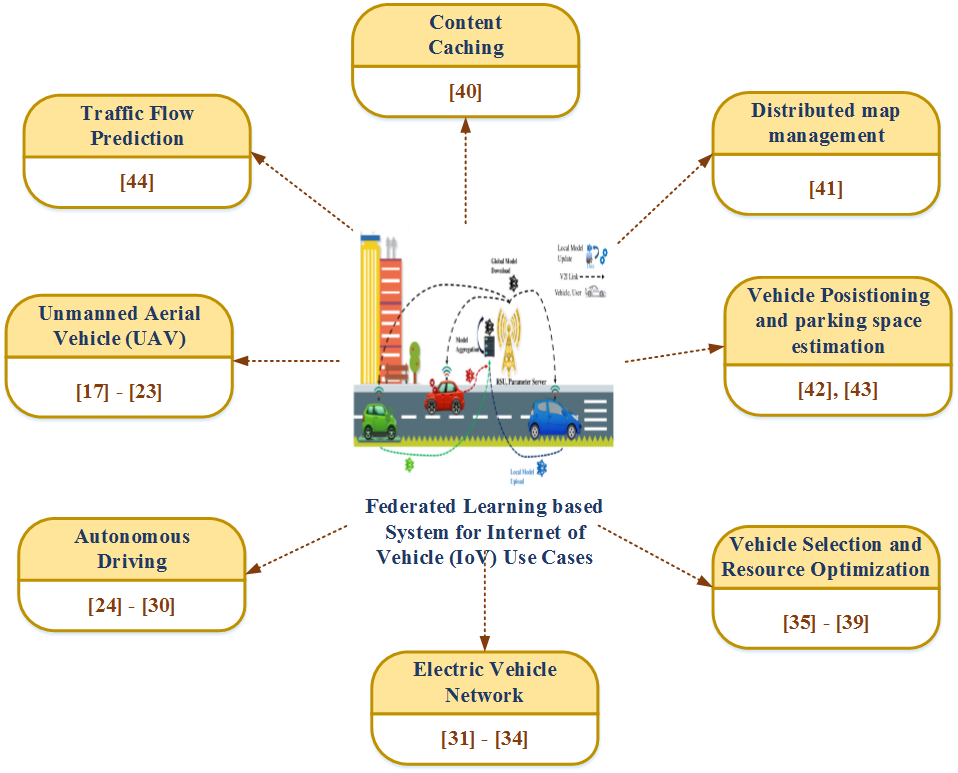}
		\label{fig:FL_Use Cases}
	\end{figure*}

	FL for developing IoV has extensively been studied in the literature. These research works aim to different application areas including: Unmanned Aerial Vehicle (UAV), Autonomous Driving, Electric Vehicle Network (EVN), Vehicle Selection and Resource Optimization, Traffic Flow Prediction (TFP), Distributed-Map Management, Content Caching, Vehicle Positioning and Parking-Space Estimation etc (summarized in Figure \ref{fig:FL_Use Cases}). \\
	
	UAVs are regularly utilized today to give information assortment and calculation offloading support in the IoV system. FL has been deployed in UAVs to deal with data privacy and security by numerous analysts. Authors in \cite{lim2021towards} proposed an FL-based approach for managing car park occupancy and traffic prediction in UAVs. To solve the limited battery life issues of mobile users, \cite{pham2021uav} proposed a scheme that applied UAV empowered wireless power transfer for assuring sustainable wireless networks based on FL. As the model parameters are exchanged continuously, the FL performance decreases with the failure of communication links and the presence of missing nodes. Therefore, to improve FL accuracy and ensure IoV communication between the server and other components,  \cite{ng2020joint} proposed to use UAVs as wireless relays. A fully decentralized FL framework named ISPW-ADMM was developed by \cite{xiao2021fully}. As air quality has significant effects on human health, accurate and timely prediction of the Air-Quality Index (AQI) is getting importance increasingly. \cite{liu2020federated} proposed an FL framework to sense the AQI of aerial-ground efficiently. An image classification method for the case of UAV-based exploration was considered by \cite{zhang2020federated}. A ground fusion center normally set in an inaccessible location played the role of coordinating multiple UAVs. On the other hand, \cite{chhikara2021federated} worked on an AQI data collection system on UAV swarm based on distributed FL.\\
	
	A vehicle network connection is normally unreliable and comparatively slow. So, it is very important to reduce communication overhead. Considering this issue, a communication efficient secure FL scheme named 'SFLEC' was proposed in \cite{li2020secure}. Another work \cite{kong2021federated} focusing on optimizing communication cost besides user privacy proposed an FL framework 'FedLPR' for recognizing license plates. On the other hand, both information leakage prevention and information transformation schemes were considered in the work \cite{lu2020federated}. A MEC server empowered synergistic data sharing among vehicles was investigated in \cite{li2021federated}. The authors combined deep Q-network and FL to ensure efficiency and security. \cite{boualouache2021federated} used FL in the 5G vehicular edge computing for passive mobile attacker detection. In the proposal presented by \cite{pokhrel2020improving}, models are trained locally by exchanging learning parameters. Again, Distributed end-edge cloud structure facilitated the two-layer FL model in the 6G environment to reduce communication overheads \cite{zhou2021two}.\\

	In \cite{mulya2020federated}, the authors proposed an electric vehicle framework for ensuring energy efficiency. To maximize the charging station profits, the authors implemented a model based on contract theory. But charging stations have issues with privacy protection mechanisms. For this privacy issue, station operators are unwilling to exchange data. Therefore, \cite{wang2021charging} proposed FL based EVN charging station recommendation system. In another work \cite{saputra2019energy}, authors developed an energy demand prediction system aiming at electric vehicular networks. Another profit-maximizing FL framework for automatically charging of electric vehicles was proposed in \cite{saputra2020federated}.\\
	
	For the cases of dynamic and complex IoV environments, an efficient solution for exchanging information according to application-specific requirements is required. Therefore, \cite{samarakoon2019distributed}, investigated the problem of resource and power allocation for communication efficient vehicular networks. Again, a perfect subset of candidate vehicle selection by the central server is essential to recover the limited bandwidth problem. Such a case was considered in the paper \cite{wang2021content}. The authors proposed a resource allocation and vehicle selection algorithm using dataset content. Similarly, another client selection scheme was proposed in \cite{bao2021edge}. In this proposal, some vehicles were assigned as edge vehicles which were later used as FL clients for training local models. Another FL-based economic vehicle selection framework for IoV was proposed in \cite{saputra2021dynamic}. In the vehicle selection framework \cite{xiao2021vehicle}, some vehicle parameters such as velocity and position of vehicles were considered for choosing the vehicles with better image quality.\\	

	Recently, there have been many advances in developing sensing accuracies and ranges for the participating vehicles using the dynamic map fusion technique. For example, in the work \cite{zhang2021distributed}, an FL framework based on dynamic map fusion is presented. The proposed scheme can gain a high-quality map in spite of uncertainty with sensing, model, numbers of objects, and missing data labels. Again, in response to the increasing content requests from various vehicles, a content caching scheme at edge nodes may be an emerging solution to decrease service latency and traffic on the network. However, a DFL-based peer-to-peer caching scheme is proposed in \cite{yu2020proactive}. Nowadays, sensing techniques and vehicle-to-infrastructure communication have been developed. Thus vehicles can communicate with neighboring landmarks for perfect positioning. An FL-based vehicle positioning system is proposed in \cite{kong2021fedvcp} which provides accurate positioning and ensures privacy. Again, a parking management system named FedParking \cite{huang2021fedparking} proposed to collaborate with parking lot operators for training an LSTM model to estimate parking space. On the other hand,  \cite{liu2020privacy} developed a TFP system considering the significance of accurate traffic prediction besides preserving privacy.

	\subsection{Blockchain Technology}
	
	  	\begin{figure*}[h!]
		\centering
		\includegraphics[scale = 0.6]{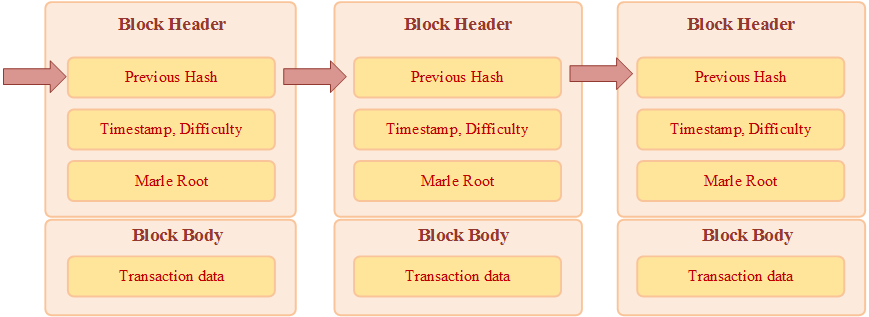}
		\label{fig:block}
	\end{figure*}
	
    BC provides a decentralized data storage environment besides transparency and security. The basic data unit of a BC system is a "block", generated by cryptography. The block records all the valid transaction information. The validity of this information is evaluated by the peers of the network. The typical structure of a 'Block' contains two sections: block header and body (Figure \ref{fig:block}). Different metadata including previous hash, timestamp, etc are stored in the block header, while the block body stores transactions data. However, BC uses different mechanisms for controlling network access. Based on these mechanisms, BC can be categorized into the following types:
	
	\begin{itemize}
		\item \textbf{Public Domain:} The consensus process is public. Anyone can participate in reading and accessing the chain and sending transactions. 
		\item \textbf{Private Domain:} Strict management on access control does not permit all the nodes to participate.
		\item \textbf{Consortium Domain:} Considered as partially decentralized as some nodes having authority can participate in the chain.
	\end{itemize}

	However, BC can be a better solution with reliability and scalability to address the privacy issues of IoV. For example, all the nodes will have equal rights in a BC-enabled IoV system. Also, they have to abide by similar obligations. As a result, though a node can not participate in the process, it will not affect the IoV operation at all. Again, being an open and transparent system, it will not require any trust establishment within the vehicles. Moreover, each and every node will participate in the maintenance work of the system. BC requires the network nodes to have a similar copy of its ledger which makes the database more reliable. Modifying a single node is totally impossible, as the data records are automatically compared. 

	\subsubsection{Blockchain Technology in IoVs}
	
\begin{figure*}[h!]
		\centering
		\includegraphics[scale = 0.5]{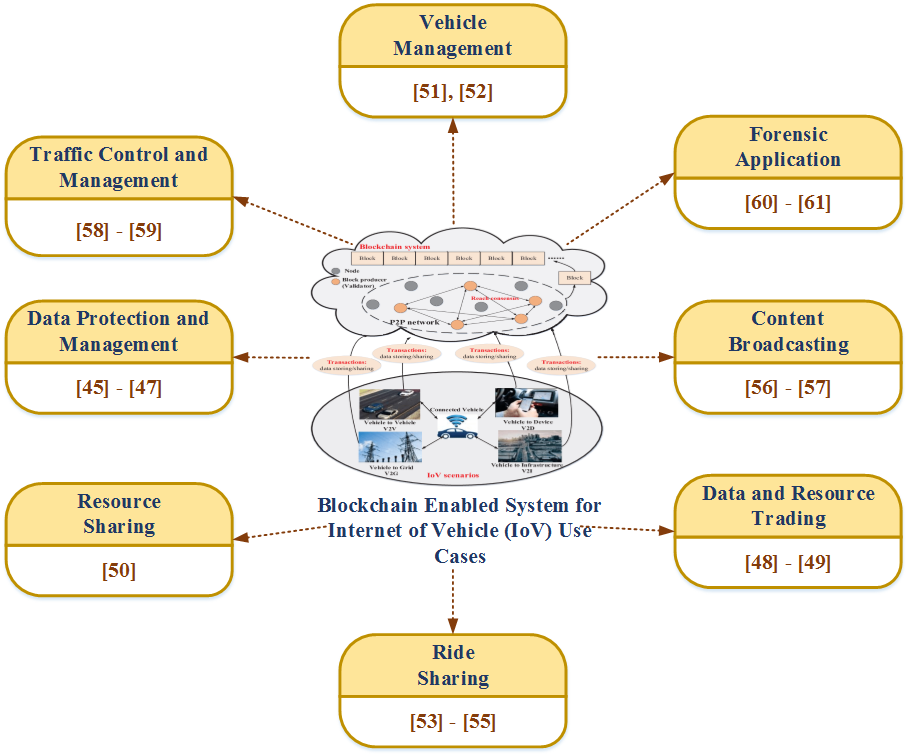}
		\label{fig:BC_Use Cases}
	\end{figure*}

	This section briefly discusses previous research works on BC-enabled IoV systems aiming to different application areas including: Data Sharing and Protection, Traffic and Vehicle Management, Resource Sharing and Trading, Ride sharing, Content Broadcasting etc (summarized in Figure \ref{fig:BC_Use Cases}).\\

	Recently, vehicular data management scenarios have adopted BC technology to address security and privacy issues and trust-building. For example, BC was deployed with an IoV network to securely manage the distributed data in the work \cite{kang2018blockchain}, while authors of the work \cite{javaid2019drivman} tried to solve the issues of trust management in the IoV system using a BC-empowered framework named 'DrivMan'. Again, authors in \cite{shi2020blockchain}  presented a  BC and cryptography-based multimedia data sharing technique that can be deployed in vehicular social networks.\\
	
	Nowadays, business opportunities have been created in the field of data and resource trading in IoV systems. This advancement has brought new security and privacy challenges also. However, BC technology is being incorporated in IoV for providing vehicular data trading. As an example, the BC-enabled framework proposed in the paper \cite{chen2019secure}, efficiently handles the data trading mechanisms in the IoV. A resource trading platform using BC is presented in \cite{li2019computing}. However, vehicles can be extensively benefited through resource sharing in the IoV platform. \cite{chai2019proof} considers the case of resource sharing using BC technology. The proposed lightweight consensus mechanism can build trust among clients and reduce the mining cost.\\

	BC technology can solve the current issues of vehicle management systems in IoVs. For example, a centralized parking system will require revealing private information like destination information during the search for free parking spaces. If BC is deployed in such management systems, vehicle drivers will be able to find parking spaces easily without losing privacy. This decentralized privacy-preserving mechanism was covered by the work \cite{al2019privacy}. BC technology can also be incorporated with the IoV system to manage vehicle platoon. Actually, vehicle platooning is useful for fuel-saving and transporting goods long-distance. Such a scheme is presented in \cite{chen2019smart} in order to establish a vehicle platoon.\\
	
	Ride-sharing services are gaining popularity nowadays due to the convenience of traveling. But ride-sharing services are mainly cloud-based and most of the services are facing some challenges such as unnecessary communication delays, risk of users’ privacy disclosure. so, researchers are focusing on integrating BC with these ride-sharing platforms. \cite{shivers2019toward} attempted to reduce the communication delay and privacy risks of the ride-sharing platform users. A new BC model named 'CoRide' for ride-sharing platform was presented in \cite{li2019coride}. \cite{li2018efficient} proposed to incorporate BC with the carpooling system. In the proposed carpooling system, users could share vehicles with proper privacy and security.\\
	
	Recently, content broadcasting in IoV scenarios is also being facilitated by BC technology. This concept is gaining popularity for advertising and promoting products besides entertainment purposes. \cite{li2019toward} showed the potential of blockchain in advertising sectors allowing the customers to promote their products. On the other hand, a different application scenario-EDM protocol is proposed in \cite{nkenyereye2020secure} which facilitates event-based messaging.\\
	
	Traffic data needs to be stored and transferred especially among the vehicles and RSUs. But, this transfer and storage process might face adversarial attacks. To reduce such risks, different approaches are proposed in the literature. \cite{cheng2019sctsc}  proposed a signal maintaining procedure for preventing adversarial attacks. The procedure was not fully centralized, but it could accelerate the efficiency of traffic control systems. Another noteworthy work \cite{ren2019intelligent} proposed an intelligent traffic system for smart transportation deploying BC technology with IoT sensors in vehicles.\\
	
	Another potential application of BC technology for autonomous vehicles is digital forensics. It can be used for analyzing traffic accidents. A vehicular forensic framework named Block4Forensic is presented in \cite{cebe2018block4forensic}. This Block4Forensic framework can be used to investigate and analyze vehicular accidental events. Software-defined network mechanism was combined with BC technology in \cite{pourvahab2019digital}. The proposed architecture could facilitate digital forensics application in the IoV system.
	
	\subsection{Convergence of FL and BC Technology}

	With a view to addressing the privacy and security issues, the centralized ML paradigm has adopted an FL framework. FL is expected to provide more advantages than the traditional decentralized learning approaches. Hypothetically, FL gives an awesome security safeguarding learning structure in IoV, a malignant vehicle may purposely alter information through poisoning attack \cite{kang2019incentive}. A lot of probable adversarial attacks including poisoning attacks and reverse engineering attacks have been reported recently in FL \cite{tolpegin2020data, zhao2021federatedreverse}. Again a self-centered vehicle may not coordinate in information assortment bringing about wrong weights of a local model. FL faces several security issues including the following:
	
	\begin{itemize}
		\item FL relies upon a local worker that can be vulnerable against cyber intrusions. If one local model is attacked, it may mislead other models and consequently, the global update is wrong.
		\cite{bonawitz2019towards}. Moreover, when various models are communicated at the same time, the central server might over-burden the network because of transmission capacity constraints \cite{elgabli2020gadmm}.
		
		\item Results predicted by the machine learning can be misleading as poisoned training samples or models can be uploaded by the malicious participants, as FL does not have the capacity to review malignant trainers. 
				
		\item FL expects that each local device contributes information assets separately. Members may not have the motivating force to participate in model training.
		
		\item FL systems may reveal the privacy and security aspects of training data, regardless of whether the training resource is put away in the local device. A few research has fostered some infer attacks based on intermediate gradients of the models \cite{bos2014private}. A malignant center may take advantage of sensitive data by Generative Adversarial Network (GAN).
	\end{itemize}
	
	However, FL has the auspicious capability to build an ITSs. On the other hand, it is crucial to alleviate any potential attacks in FL. So, BC technology is being utilized with FL to give a decentralized arrangement, for controlling incentives and guaranteeing security and protection in a dependable way \cite{lu2020low}. BC technology consists of many immutable blocks based on a set of rules named consensus which helps to form a distributed ledger. BC technology contributes to both FL and IoV for its decentralized nature \cite{pokhrel2020federated}. BC has a unique transaction verification process. So, it can be used for validating local training models in FL \cite{lu2020communication}. BC is utilized to empower secretly and safely sharing of data, and give a completely auditable log of data. BC empowered FL structures to empower a protected and dependable collaborative learning for different ITS features like traffic monitoring, intrusion detection, autonomous driving, driving in different weather conditions, perfect routing, etc.	

\section{Hybrid BC-based FL Model for IoVs}
\label{BFL_IOV}

	Architectures and frameworks that integrates both BC and IoV technologies together will be presented in the following section.
	
	FL appears to be a powerful technique for privacy preservation. However, FL also faces vulnerability to poisoning, reverse engineering, and other types of attacks. Security of the clients can be uncovered by the intruders if model updates received from the clients can be manipulated. Also, gaining control over the aggregating server, an intruder can acquire extensive information on historical parameters and global model structure. This information can then be applied to reverse engineering for uncovering the client's security. Moreover, if the intruder can control the aggregator server, he/she may produce the global model parameters which may influence the local model parameters. However, regardless of whether the intruder has control over the server or local devices, both of the model's parameters may, in any case, be forged while on the way among client and server. In the event that the attacker deals with the devices and has access to the local models or data, certainly it will influence the accuracy of the global model. Because of the promising capability of FL, especially for building an ITS and the requirement for relieving potential attacks, various BC-enabled FL schemes for IoV have been proposed. In this section, we describe the literature works on hybrid Blockchain-enabled federated learning frameworks for intelligent transportation systems in terms of used methods, experimental parameters, dataset, libraries or platform for implementation, outcome, the main focus, and application scenarios. The summary table of literary works of the Hybrid BC-based FL model for IoV is provided in Table \ref{tab:BC FL in IoV}.
	
	\subsection{Experimental Works on Hybrid BC-based FL Models} 
	
	Many research works experimented with their proposed system based on the popular MNIST/CIFAR-10 dataset. \cite{chai2020hierarchical} proposed a hierarchical BC-enabled FL framework focusing on knowledge sharing. They focused on fulfilling the privacy requirements of IoVs for large-scale vehicular networks. Authors modeled knowledge sharing as a trading market process formulated as a multi-player game. Both the vehicles and RSUs can collect surrounding data. For attracting participants in knowledge sharing, BC was adopted to record learning results. In most of the scenarios vehicles collect data in the diverse region as driving routes also varies. Considering such cases, the data collection and knowledge sharing process was enabled through a hierarchical BC ledger. However, the proposed ADMM-based algorithm enhanced accuracy over 10\% compared to conventional FL algorithms.\\
	
	Another work \cite{lu2020blockchain} simulated on MNIST dataset integrated blockchain and FL-based directed acyclic graph to build a hybrid BC architecture. They attempted to mitigate transmission load and solve privacy issues of providers. Deep Reinforcement Learning was adopted in an asynchronous FL scheme for node selection to increase efficiency. However, to simulate the proposed scenario, 1500 by 1000 meter area was defined on New York City map with Uber pickups datasets as trace-points of vehicles. \\
	
	Another paper \cite{wang2021blockchain} proposed an IoV system to integrate MEC with BC-empowered FL framework. Specifically, they introduced differential privacy techniques to prevent privacy concerns while the honesty of participants is ensured by a malicious updates remove algorithm based on the self-reliability filter. Moreover, a double aggregation frame is proposed, to guarantee the communication overhead and ensure the quality of model training. Simulation results on the MNIST dataset show that it effectively defends the backdoor attack and remains stable with a 9.54\% attack success rate. \\

	Security challenges of the mistrustful centralized trading model are addressed in the work \cite{zou2021reputation}. The authors proposed a BC-enabled knowledge trading scheme. An approved market agency organizes the trading rapidly. Considering the conflict of knowledge providers, they modeled the pricing mechanism as a non-cooperative game. Simulation of proposed knowledge trading scheme on MNIST dataset claims to improve the knowledge accuracy up to 18\%.\\
	
	\cite{wang2020learning} proposed a framework called SFAC-secured FL scheme in UAVs assisted mobile crowd-sensing. At first, they introduced three attacks and investigated the corresponding defenses. Then Secondly they implemented a decentralized blockchain network that records and traces UAVs' contributions. On the other hand, they devised an LDP mechanism for securely sharing local models with better aggregation accuracy. The simulation of the proposed SFAC framework is carried out in python programming language on the MNIST dataset. The convergence time (time slot) for the SFAC framework with four base stations was 2180.
	
	\subsection{Theoretical Works on Hybrid BC-based FL Models}

	Some research works focus on the theoretical aspects of such models. Such literature includes \cite{pokhrel2020decentralized}, where authors designed an autonomous transport system that can preserve the privacy of the vehicles data and vehicles can efficiently communicate with each other. Updated ML models residing on the local vehicles are verified and exchanged based on a distributed fashion. The local models do not require any centralized training data. Again, the application of the BC consensus scheme eliminates the necessity of any central coordination. A mathematical framework based on the renewal reward approach was developed. Different network and BFL parameters including frame and block sizes, block arrival rate, maximum re-transmission were featured to assess their impact on the performance of the system. In another work \cite{pokhrel2020federated}, the same author designed an autonomous BC-based FL scheme. They developed a mathematical framework to feature the mentioned parameters. They proposed a uniform random vehicle-miner and a comprehensive mathematical framework to feature end-to-end delay analysis. Another theoretical proposal for protecting disaster response systems through a drone-assisted BC-enabled FL at 6G edge is carried out by \cite{pokhrel2020federatedUAV}. Their main focus was on reducing latency in BC and consumption of energy in the drone network. It is required to maintain low communication delay in wireless drone networks to reduce the occurrence of forking events in the BC technology. In this proposal, the author finds out the probability of such events to anticipate system uncertainty towards further energy loss. \\
	
	Another noteworthy proposal on pedestrian safety in autonomous driving systems integrates BC and FL technology in \cite{kansra2022blockfits}. The authors proposed a model training paradigm named 'BlockFITS' consisting of two-way communication, Vehicle-to-BC (V2B) and BC-to-Vehicle (B2V). To provide incentives to the participants, a data augmentation scheme was operated with cooperative training. In another paper \cite{majeed2019flchain}, authors proposed a BC-based model named 'FLchain' aiming to enhance FL security. A concept of the channel was used in their proposed work for handling multiple servers containing global models. After each global iteration, local model parameters are stored within the channel-specific ledger as blocks.\\
	
	\cite{ghimire2021secure} presented a theoretical framework scenario of BC-empowered FL for privacy-preserving and verifiable FL for IoV to provide secure and trustworthy ITS services. The proposed approach hypothetically provides assurances to give a powerful and secure FL environment. The registration process of the vehicle through authorities like the DMVs limits the security dangers fundamentally. Moreover, BC empowers members to utilize pseudonyms as BC identity to conceal the genuine identity. On the other hand, any chance of adversarial attacks is eliminated through the use of pseudonyms and model parameter exchange through BC networks. Again, the miner vehicles verify the federation aggregation process to mitigate a single point failure for malicious activity from the server-side. Though it is hard to keep clients from acting maliciously (either itself or through ill-disposed control), BC hubs run IDS that can dissect the parameters’ history to recognize any adversaries and go to remedial steps. \\
	
    	
	
	\subsection{Real-time Works on Hybrid BC-based FL Models}

	Besides theoretical proposals, some works experimented with their proposals using real datasets. Vehicular organizations built by interconnected vehicles are exposed to adversarial attacks because of the extended use of the software. There has been critical advancement in distinguishing pernicious attacks utilizing machine learning techniques. \cite{liu2021blockchain} proposes an autonomous driving structure in which both model training efficiency and security of model sharing are incorporated. There may have some cases when an edge vehicle will try to corrupt the global model by uploading malicious updates. To prevent such uploading of poor-quality edge model parameters, authors designed a mechanism for tracking data and selecting models for RSU. RSU divides edge models into different groups and aggregates edge model parameters for each group. Comparing the aggregation accuracy of the groups, RSU finds out the poisonous model parameters and avoid poisoning attacks effectively. However, for implementing the experiment, third-party libraries Pytorch and Syft and a publicly available KDDCup99 \cite{bolon2011feature} dataset (US Air Force LAN data) are used. Different measurement of the accuracy of the model (Best accuracy is 96\% for dataset size 10,000 and epoch number 40) is achieved in accordance with data set sizes and number of epochs.\\
	
	\cite{li2021privacy} introduced a FL framework for autonomous vehicles. Like other FL frameworks, they preserved data in local vehicles while sharing only the model parameters with the assistance of the MEC server. Instead of only honest MEC servers and honest vehicles, the authors also considered the case of malicious vehicles and MEC servers. In the case of honest-but-curious MEC servers and malicious vehicles, a traceable privacy-preserving scheme based on identity was proposed. This scheme was adopted for protecting vehicular message privacy. On the other hand, for the case of semi-honest MEC servers and malicious vehicles, their proposed privacy scheme was based on anonymous identity.\\
	
	In \cite{shen2020blockchain}, authors proposed a blockchain-enabled distributed federal learning approach to accurately predict the rail transit passenger flow. Instead of a trusted central server, it performs distributed machine learning. The management of the whole federal learning is realized by the blockchain smart contract. Each station stores corresponding passenger flow data. Those data were converted into time-series data for training the LSTM neural network directly. Each station uses its local
	
	\onecolumn	
	\begin{longtable}{|c|c|p{4cm}|p{2cm}|p{2cm}|p{3cm}|p{2cm}|}
		\caption{Summary of literature works of Hybrid BC-based FL model for IoVs in terms of used methods, experimental parameters, dataset, libraries or platform for implementation, outcome, main focus	and application scenarios.}
		\label{tab:BC FL in IoV} \\ 
		\hline
		\textbf{Ref.} & \textbf{Year} & \textbf{Methods} & \textbf{Experimental Parameters / Dataset} & \textbf{Libraries / Platform / Implementation} & \textbf{Outcome} & \textbf{Main Focus} \\ 
		\hline
		\endhead
		
		\cite{aloqaily2021energy} & 2021 & Deploy both UAVs and UGVs in vehicular environments to continuously provide connectivity, UGVs act also as moving charging stations & traffic data set extracted from U.S. Traffic Fatality Records & Python and the scikit-learn ML library, NS-3, OMNET++, OverSim framework & Energy consumption reduction: 89\%, average network coverage ratio: 97\%, packet delivery success rate: 90\% & UAV energy enhancements\\ 
		\hline
		
		\cite{pokhrel2020federatedUAV} & 2020 &  Finds out the probability forking events to assess the uncertainty of the system towards further energy wastage & Theoretical proposal, Not Implemented & Theoretical proposal, Not Implemented & Theoretical proposal, Not Implemented & blockchain latency and energy consumption \\ 
		\hline
		
		\cite{wang2020learning} & 2021 & introduced three attacks,  investigated the corresponding defences, devised an LDP mechanism-based privacy preserving local model sharing algorithm & MNIST &The CNN model is adopted for model training, Python & Convergence time (time slot): 2180 & High-quality model sharing and ensure privacy protection for UAVs \\ \hline 
		
		\cite{joshi2020toward} & 2021 & Blockchain enabled rain drop optimization (RFO) algorithm- RAOC-B & Simulation of urban mobility (SUMO) generated Data & Network Simulator-2, SUMO & End to end delay = 0.3 s for 100 nodes; maximum Packet delivery ratio = 0.9; Highest throughput = 94\% & packet delivery ratio, end to end delay, throughput, and cluster  size \\ 
		\hline
		
		\cite{ayaz2021blockchain}      & 2020   & Proof-of-Federated-Learning (PoFL) based message dissemination   & OMNeT++ simulator generated data   & OMNeT++, Python, SUMO and VeINS  &65.2\% faster and at least 8.2\% more efficient in message dissemination approach.  &message dissemination         \\ \hline
		
		\cite{pokhrel2020decentralized} & 2020  & Updated machine learning models residing on the local vehicles are verified and exchanged based on a distributed fashion   & Theoretical proposal, Not Implemented & Theoretical proposal, Not Implemented&  Theoretical proposal, Not Implemented   & Efficient communication of autonomous vehicles \\ \hline
		
		\cite{chai2020hierarchical}  & 2020   & ADMM-based algorithm      & MNIST,	CIFAR10  & No info about Library/ Environment except simulation parameters &10\% more accuracy enhancement over conventional FL algorithms  & Knowledge Sharing in Internet of Vehicles   \\ \hline
		
		\cite{lu2020blockchain}  & 2020  & DRL-based node selection algorithm. &MNIST   &  matplotlib basemap toolkit & Improved accuracy more than 90\%     & Secure Data Sharing in Internet of Vehicles   \\ \hline
		
		\cite{wang2021blockchain}  & 2021 & privacy-preserving FL framework, named BMFL, mobile edge computing (MEC), differential privacy (DP) &MNIST &IoV system with 50 devices, 5 MEC servers and a cloud server & effectively defends the backdoor attack and remains stable with 9.54\% attack success rate  &  reducing cloud communication overhead and ensuring the quality of model training  \\ \hline

		\cite{otoum2020blockchain} & 2020  &  decentralization mechanism was considered in VNet systems & MATLAB/ Simulink generated data  & MATLAB/ Simulink; Contiki operating system; python & 97\% accuracy & Accuracy in the  Vehicular Network (VNet) environment \\ \hline
		
		\cite{kansra2022blockfits}  & 2021  & Vehicle-to-BlockChain-to-Vehicle (V2B2V) federated learning enabled model training paradigm for ITS entities
		& Theoritical proposal, Not Implemented  & Theoritical proposal, Not Implemented &Theoritical proposal, Not Implemented & Autonomous driving \\ \hline
		
		\cite{pokhrel2020federated}  & 2020  &BFL enables oVML without any centralized training data or coordination by utilizing the consensus mechanism of the blockchain. & Theoritical proposal, Not Implemented  & Theoritical proposal, Not Implemented &  Theoritical proposal, Not Implemented  & Autonomous Vehicles  \\ \hline
		
		\cite{majeed2019flchain}   & 2019  & Concept of channel was used on their proposed work for handling multiple servers containing global models  &  Theoritical proposal, Not Implemented   & Theoritical proposal, Not Implemented  & Theoritical proposal, Not Implemented 	& Enhancing security of Federated Learning (FL) \\ \hline
		
		\cite{li2021privacy}      & 2021  &improved Dijk-Gentry-Halevi-Vaikutanathan (DGHV) algorithm & Driving and road condition data of Rancho Palos Verdes and San Pedro California  &autonomous driving simulation in Python with the real-world data  &reduce around 73.7\% training loss	& Autonomous car  \\ \hline
		
		\cite{zou2021reputation} & 2021   &CNN model is used as the local training model  & MNIST   & Numerical simulation   &  improves the accuracy of knowledge up to 18\%  & Knowledge Trading  \\ \hline
		
		\cite{liu2020privacy} & 2020 & Federated Learning-based	Gated Recurrent Unit neural network algorithm (FedGRU)    &Caltrans Performance Measurement System (PeMS) Dataset    &5 time steps, 2 hidden layer, hidden units 50, 50  &Accuracy = 90.96\%     &Traffic Flow Prediction  \\ \hline
		
		\cite{peng2021bflp} & 2021  &construct a lightweight encryption algorithm called CPC   &No information found on dataset  &TOSSIM simulator;
		TinyOS system & rate of predicting road condition = 84.25\%,	Required time = 4000 ms,cost = 13000 bytes   & predicting road conditions   \\ \hline
		
		\cite{qi2021privacy} & 2021  &  neural network called GRU, differential privacy, central server is replaced by a set of trusted consensus nodes  & Caltrans performance
		measurement system dataset & PySyft for FL framework; consortium blockchain  &MAE = 7.96; MSE = 101.49; RMSE = 11.04  &Traffic Flow Prediction \\ \hline
		
		\cite{liu2021blockchain} & 2021  & Comparing the aggregation accuracy of different groups, RSU infer the poisonous model parameters & KDDCup9 \cite{bolon2011feature} & Pytorch,  Syft and go language & Accuracy = 96\% for Data   Size  = 10000 and Epoch = 40 & Handling poisoning attacks\\ 	\hline
		
		\cite{ghimire2021secure} & 2021 &only restricted parties  are registered;	aggregation is always verified by the miner vehicles to prevent malicious activity from
		the server side  &  Theoretical proposal, Not Implemented & Theoretical proposal, Not Implemented & Theoretical proposal, Not Implemented  &reverse engineering attacks elimination,intrusion detection system \\ \hline
		
		\cite{shen2020blockchain} & 2020  & long short-term memory (LSTM) network has been used as the supervised learning model  &passenger flow data of Beijing Metro &Linux, EOS	for blockchain and Node.js for test script. &  Optimal prediction of LSTM;  & Secure Railway passenger flow prediction model \\ \hline
		
		\cite{hua2020blockchain} & 2020 &  Support vector machine (SVM) has been used as the supervised learning model  & train running data from Xiaojue Station of Shuohuang Railway to West Station of Dingzhou & svm model based on the mixed kernel function & Accuracy with federated learning is 94.21\%. & Intelligent control model for heavy haul trains\\  \hline
		
	\end{longtable}
	\twocolumn

	\noindent data to train an LSTM model and calculate the gradient values and error rate of the model. Then the stations upload the gradients and error rate onto the BC as a transaction. The model containing the lowest error rate is shared with other stations, which then use this model for training themselves. As the station which can not precisely predict using its local data is assumed to contain more information than others. Then, this station trains its local data using the first best model. Above mentioned steps are continued iteratively until the error rate reaches a predefined threshold.\\
	
	Authors of the work \cite{shen2020blockchain} also proposed another BC-enabled FL scheme to perform asynchronous and collaborative ML among distributed rail agents in \cite{hua2020blockchain}. Like the previous work, instead of a trusted central server, it performs distributed ML. The entire FL management is realized by BC smart contract. To simulate the experiment, the authors collected historical driving data from rail systems. The traditional SVM model is optimized by penalizing both majority classes and minority classes to deal with the imbalanced data.\\
	
	In the work \cite{qi2021privacy},  authors used a lightweight neural network called GRU for processing traffic data. The central server is replaced by a set of trusted nodes who is responsible for managing all the local model updates. These nodes are also called miners to verify the model updates from the local vehicles, which are then stored on the blockchain. For ensuring the privacy of the participating client vehicles, a differential privacy mechanism is applied via the noise-adding process with the vehicle uploaded local models. Thus it provides location information privacy and prevents membership inference attacks to collect vehicle information. The work is evaluated on Caltrans performance measurement system \cite{choe2002freeway}  dataset and achieves the better result (MAE is 7.96, MSE is 101.49 and RMSE is 11.04).\\
	
	\cite{liu2020privacy} introduced an FL scheme based on neural network algorithm named 'FedGRU' for TFPs. The proposed method updates the global model through an improved federated averaging algorithm. It reduces communication overhead while model parameters are transmitted. In order to improve scalability, a joint announcement protocol is designed. On the other hand, an ensemble clustering-based scheme is proposed which groups the vehicles into clusters. The work is evaluated on Caltrans performance measurement system \cite{choe2002freeway} dataset. The FedGRU can produce predictions with a better result (MAE is 7.96, MSE is 101.49 and RMSE is 11.04).\\

	In the work \cite{aloqaily2021energy}, the authors designed a model that deployed both UAVs and UGVs to reduce the power consumption of aerial vehicles. For ensuring longer and prolonged service continuity for the UAVs, ground vehicles

	\noindent were considered as charging stations. An optimization solution was carried out by the authors to ensure a longer power range for vehicular devices. On the other hand, a blockchain-enabled FL mechanism is deployed to provide data privacy. Different experimental environments were set up to simulate the proposed solution including Python and the scikit-learn ML library for testing the power management issue, NS-3 for testing the connectivity issue, OMNET++ for testing the service provisioning process. The proposed strategy decreases the overall UAV energy consumption by 89\% compared to the non-cooperative solution. On the other hand, the average network coverage ratio is about 97\%, where the packet delivery success rate is 90\%.
	
	\subsection{Synthetical works on Hybrid BC-based FL Models} 
	
	A good number of research works consider simulators like Matlab Simulink, OMNET++, Network Simulator-2 (NS-2), etc for generating synthetic datasets. \cite{otoum2020blockchain} proposed an integrated BC-based FL scheme to serve VNet architecture. To ensure trustworthiness and reduce delay decentralization mechanism was considered in VNet systems. The proposed model was simulated using the MATLAB/Simulink package in Contiki operating system. Testing the proposed hybrid model on 20 nodes achieved 97\% accuracy in the IoV environment.\\
	
	A noteworthy work on vehicular message dissemination is proposed in \cite{ayaz2021blockchain}.  Actually to ensure road safety message exchange among vehicles has an important role. Usually, emergency message dissemination is performed through broadcasting. But, increasing vehicle density and mobility are leading to challenges in message dissemination (e.g. broadcasting storm and low probability of packet reception). The authors in this work proposed a BC-based FL solution for message dissemination. Especially. their proposed Proof-of-Federated-Learning (PoFL) consensus attracts more vehicles to compete for model training which leads to a more accurate model. The proposed method is claimed to outperform the other blockchain solutions for message dissemination to reduce 65.2\% time delay in consensus and to improve message delivery rate (at least 8.2\%). Also, the authors analyzed the economic model for incentivizing vehicles taking part in federated learning and message dissemination using the Stackelberg game model.\\

    Recently, vehicular ad hoc networks also called 'VANET' have emerged as a key part of IoV systems. \cite{joshi2020toward} introduced a cluster-based vehicular ad hoc network deploying BC technology, that can efficiently preserve user privacy during data transmission. Using 'Rainfall Optimization Algorithm' the vehicles in the VANET are clustered. ROA creates different clusters of vehicles, where a vehicle works as Cluster Head (CH) for every group. This balances the load effectively and creates less congestion in the network. They have named this ROA-based technique merged with BC-enabled data transmission as ROAC-B technique. ROAC-B primarily groups the vehicles and communication that occur through BC technology. Various grid sizes are analyzed in terms of end-to-end delay (0.3 s for 100 nodes), packet delivery ratio (0.9), and throughput (94\% at 100 seconds of simulation). To simulate the framework, Network Simulator-2 is used which needs a 'SUMO' simulator for generating the mobility of vehicles.\\

	\cite{peng2021bflp} proposed a framework named BFLP, which allows local model training tasks without sharing raw data, and it can select the most perfect FL method according to the application scenarios. Considering the poor computing capability of vehicles, a lightweight encryption algorithm called CPC is constructed to preserve privacy. The authors conducted experiments in obstacle detection and traffic forecast scenarios. BC technology was applied at the bottom layer to secure the data transmission and protect users’ privacy. They used the TOSSIM simulator to design the vehicle's base station model. execution time of CPC was evaluated on TinyOS. The experiment focused on predicting road conditions and achieved a rate of predicting road conditions is 84.25\%. Required time and cost (space) for CPC algorithm is 4000 ms and 13000 bytes respectively.\\
	
\section{Some Application Scenarios of BC-Enabled FL Frameworks for IoVs}
\label{BFL_AS}   
	
\begin{figure*}[!h]
		\centering
		\includegraphics[scale = 0.7]{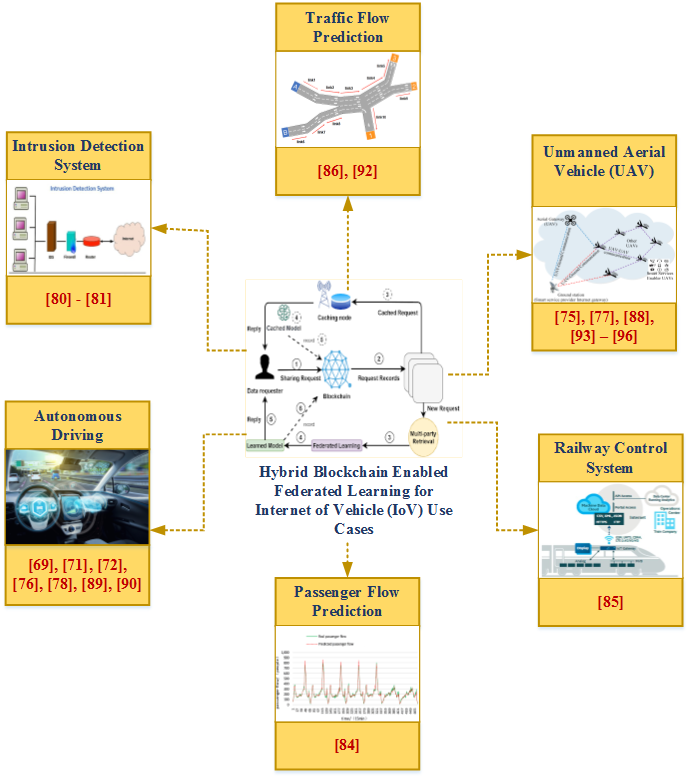}
		\label{fig:HBCFL_Use Cases}
	\end{figure*}

	After going through the works presented in the literature on how BC-enabled FL technology can be utilized in IoV application scenarios, we determine that BC-enabled FL frameworks are anticipated to cope with a number of IoV applications and services including Unmanned Aerial Vehicle (UAV), Traffic Flow Prediction (TFP), Intrusion Detection System (IDS), Passenger Flow Prediction (PFP), Railway Control System (RCS), and Autonomous Driving as illustrated in Fig. \ref{fig:HBCFL_Use Cases}. In the following subsection, we review and investigate the incorporation of BC-enabled FL frameworks and such IoV application scenarios.
	
	\subsection{Unmanned Aerial Vehicle}
	
	As modern traffic networks are getting complex day by day, aerial platforms are gaining importance increasingly beyond ground data sources. Specially for data collection and supporting computation offloading in the IoV environment, UAVs are being used commonly nowadays. Moreover, cost-effectiveness, high mobility, and flexible deployment are featured by UAVs besides more effective coverage \cite{zhou2014efficient}. The image capturing for car parking management \cite{zhou2017car}, traffic monitoring \cite{elloumi2018monitoring, coifman2006roadway}, data aggregation from vehicles and RSUs can also be benefited through the deployment of UAVs. However, some fundamental challenges should be taken into consideration to get benefited from federated learning-based UAV systems. For example, collaborative participation of UAVs can not be ensured because of resource constraints and selfishness. Also, there are probabilities of low-quality sensor data and insufficient training samples. Again, the vulnerability in centralized model aggregators can lead to whole process failure as evidence for rewards of participants can be tempered. Moreover, all of the privacy concerns can not be eliminated by federated learning. Thus, several proposals are found on secured and privacy-preserving UAV schemes to promote FL collaboration. For example, \cite{aloqaily2021energy} deployed both unmanned aerial and ground vehicle systems in the IoV system to continuously provide connectivity where UGVs acted as moving charging stations. Based on traffic data set extracted from U.S. Traffic Fatality Records, they focused on UAV energy enhancements. Another work \cite{wang2020learning} attempted to ensure high-quality model sharing and privacy protection for UAVs. An exceptional proposal is found in \cite{pokhrel2020federatedUAV}, where the author proposed a disaster management system assisted by drones. In this work, the main attempt was to minimize latency in blockchain and consumption of energy in the network of drones.

	\subsection{Intrusion Detection System}
	
	Cyber-attacks are becoming more common as software and wireless interfaces become more widely used, as well as vehicle networks and intelligent transportation infrastructure. In response to these attacks, Intrusion Detection Systems (IDSs) can be deployed to detect malicious attack traffic. However, offload the training model to distributed terminal devices, reduce resource utilization of the central server and preserve security and privacy of the aggregation model, several schemes are proposed. \cite{liu2021blockchain} proposed an IDS that can handle poisoning attacks with an accuracy rate of 96\%. Another IDS \cite{ghimire2021secure} proposed a theoretical scheme to eliminate the probable chance of spoofing, forging, and reverse engineering attacks elimination.
	
	\subsection{Traffic Flow Prediction}
	
	Traffic Flow Prediction (TFP) is considered a necessary component for the effective implementation of ITS subsystems, particularly sophisticated traveler information, online car-hailing, and traffic control systems. However, existing TFP mechanisms based on FL frameworks on a centralized model still suffer from severe security issues, such as a single point of failure. To address this problem, BC is combined with the FL model to provide a decentralized, dependable, and secure model that does not require a centralized model coordinator. For instance, \cite{peng2021bflp} constructed a lightweight encryption algorithm called CPC for predicting traffic conditions. Their rate of predicting road conditions was 84.25\% that required 4000 ms time. On the other hand, \cite{qi2021privacy} worked on traffic prediction using a neural network called GRU and differential privacy with an RMSE value of 11.04\%.
	
	\subsection{Railway Control System}
	
	Due to the extended train marshaling and complex line conditions in heavy-haul-rail systems, operating modes frequently change during train passage. Safe operations of trains will be severely affected by train decoupling because of longitudinal impact force to trains caused by improper traction or braking operation. So, manual control should be replaced with intelligent control systems in heavy-haul-rail systems. Traditional machine learning-based intelligent control mechanisms, on the other hand, suffer from insufficient data. Again, data from multiple train lines or operators cannot be communicated directly in the FL model due to a lack of effective incentives and trust. An intelligent control mechanism has been obtained by a fusion algorithm to achieve the intelligent control system of traction/electric brakes of heavy-haul trains \cite{hua2020blockchain}. For simulation, they used train running data from Shuohuang's Xiaojue station to Dingzhou's West station. The overall prediction accuracy with the FL model is 94.21\%.
	
	\subsection{Passenger Flow Prediction}
	
	The ability to accurately predict passenger flow can aid in optimizing the vehicle management plan and increasing operational efficiency. However, only one work is found on passenger flow prediction of urban rail transit system \cite{shen2020blockchain}.  From March 1, 2014, to March 31, 2014, full-day passenger flow data from each line of the Beijing metro were utilized to train distributed Long Short-Term Memory (LSTM) model in order to create a secure railway passenger flow prediction model.	
	
	\subsection{Autonomous Driving}
	
	With limited processing resources and datasets, vehicles are unable to train a high-accuracy autonomous driving model in a low-latency manner. Many studies focus on Multi-access Edge Computing (MEC) server-assisted autonomous driving systems to tackle this challenge. Autonomous vehicles could train a more accurate model in a shorter amount of time with the help of powerful MEC servers. For example, \cite{joshi2020toward} developed a BC-enabled Rain Drop Optimization (RFO) algorithm to increase Packet Delivery Ratio (PDR), throughput, and decrease End-to-End (ETE) delay of autonomous vehicles. Another work \cite{ayaz2021blockchain} used Proof-of-Federated Learning (PoFL) to speed up and improve the efficiency of autonomous vehicle message distribution by 65.2\%. To improve knowledge sharing among autonomous vehicles \cite{chai2020hierarchical} proposed Alternating Direction Method of Multipliers (ADMM) based algorithm. BC-empowered asynchronous FL model for secure data sharing among different channels in the IoV system was proposed by \cite{lu2020blockchain}. \cite{kansra2022blockfits} proposed Vehicle-Blockchain-Vehicle (V2B2V) FL-enabled model for autonomous driving system, pedestrian safety, and vehicular object detection. Some theoretical proposals \cite{pokhrel2020decentralized, pokhrel2020federated} are also found in focusing on efficient communication of autonomous vehicles.
	
\section{Challenges, Solutions, and Future Research Directions}
\label{challenges}

    A review of BC-enabled FL frameworks is presented in this research in order to develop fully safe and robust IoV systems. In this domain, a total of twenty-two (22) research papers were considered. Some of the studies simply describe theoretical schemes, while others additionally show simulation results. However, proposed BC-enabled FL frameworks in IoV systems are still in their early stages. To fully realize the integration's potential in this domain, we have listed the major open research challenges, solutions, and possible research directions. 
	
    \begin{enumerate} 
	
	    \item Most of the research indicates that more participating vehicles result in higher performance of the final model. Thus, attracting more vehicles is a key challenge for the BC-enabled FL model of IoV systems. Devising a perfect reward mechanism for user devices can be a proper solution to this problem.
		
	    \item For sending every transaction to the BC network, user devices rely on the integrity of their associated edge devices or terminal devices.
		
    	\item Some of the research works just proposed theoretical schemes only. Without experiments or simulations, it is difficult to analyze such proposals. On the other hand, most of the experiments were based on simulator-generated data. Deploying real datasets will increase the acceptability of the proposed schemes.
		
    	\item Forking may turn out to be  major challenge in real cases. For example, if a vehicle can not access the most recent block due to unavoidable network delays, it may create a different branch chain. But it is not possible to exist more than one chain. Ultimately, blocks of other chains will be dropped. Thus, such forking event will reduce system performance. 
		
    	\item Though some researchers used consensus algorithms of BC technology for avoiding poison attacks, these also face some security flaws. For example, the misdeed miners who require maximum computing capacity, may lead to BC forking events.
		
    	\item Some studies designed and implemented new consensus algorithms. Though these algorithms reduce power consumption, but the security is not improved. So, secured consensus algorithm should be developed in public blockchain for BC-enabled FL frameworks for IoV systems is necessary. However, private blockchain or consortium technology is guaranteed to provide more security to IoV systems.
		
    	\item The security of BC-enabled FL frameworks is improved, but there is no privacy protection. As a result, additional privacy-preserving approaches such as differential privacy should be included. Differential privacy is able to prevent attackers from inferring privacy information by extracting the learned model. So, it is recommended to integrate more privacy-preserving methods with IoV systems.		
		
	    \item Different data quality-driven metrics can be added to improve the incentive techniques. For example, the client who provides dataset with the highest diversity or any trend may be considered for providing incentives. 
		
    	\item Some common performance measurement techniques have been used by most of the research works to test their proposed method. More tests should be considered including accuracy rate, latency, throughput, required time, lifetime reduction, energy consumption, cost, network coverage, packet delivery success rate, packet delivery ratio, throughput, end to end delay, etc.
		
    	\item Different FL algorithms including ensemble learning techniques should be considered with different datasets.
		
	    \item Very few proposals are found to test their model with different types of attacks such as poisoning attacks, backdoor attacks, sign-flipping attacks, same-value attacks, and reverse engineering methods. So, the actual impact of incorporating blockchain with the FL technique is not well understood.
		
	    \item Though many issues of FL technique can be improved by BC technology, vehicle heterogeneity, systems heterogeneity, and statistical heterogeneity may arise as critical issues which should be resolved to ensure the efficiency of BC-enabled FL frameworks for IoV systems.
		
	    \item Overhead and transaction throughput of the BC- enabled FL frameworks should be assessed in a practical IoV environment.
	    
	    \item Though some proposals discussed message dissemination and relay selection. However, it can further be improved by including cross-layer information in the dataset, obtained from physical and MAC layers.
	    
    \end{enumerate}
	
\section{Conclusion}
\label{Conclusion}
	
    Automobile manufacturers are focused on developing completely autonomous vehicles that will provide enough security. IoV is the core technology behind any ITS and autonomous driving, and it's being used to tackle current traffic challenges like traffic prediction and traffic management applications. To construct a productive and powerful ITS, a learning system should be set up, which does not just give street safety and other traffic-related administrations only, additionally has the option to distinguish any sort of inconsistencies and interruption and take remedial measures. To adapt to the rise in probable privacy and security issues, the centralized ML paradigm has been transitioned in FL technology. Though FL gives an awesome security safeguarding learning structure, it generally depends on a central aggregator. Moreover, it needs a supportable economic model to boost mobile devices for their contributions and adversary attacks prevention. Motivated by the auspicious capability of FL for building an ITS and the requirement for alleviating any potential attacks in FL, the BC technology is being utilized with FL to give a decentralized arrangement, for controlling incentives and guaranteeing security and protection in a dependable way. In this paper, we have presented a comprehensive survey on BC-enabled FL frameworks for the IoV systems. First, we have discussed FL technology from the perspective of network topology, data partition, data availability, aggregation algorithm, and open source frameworks. A brief review of the FL-based IoV system has also been discussed. Next, we briefly discussed BC technology and some noteworthy works on BC-based IoV systems. Works on the integration of FL and BC in the IoV system on the basis of proposed methods, dataset, outcome, key-focus, and platform are presented in this paper. Also, some use cases of such hybrid frameworks including UAV, TFP, IDS, PFP, railway, and autonomous driving have also been discussed. We then concluded by highlighting challenges, open issues, and future research directions in BC-enabled FL frameworks for the IoV systems. 

\bibliographystyle{unsrt}
\bibliography{access}

\begin{IEEEbiography}[{\includegraphics[width=1in,height=1.25in,clip,keepaspectratio]{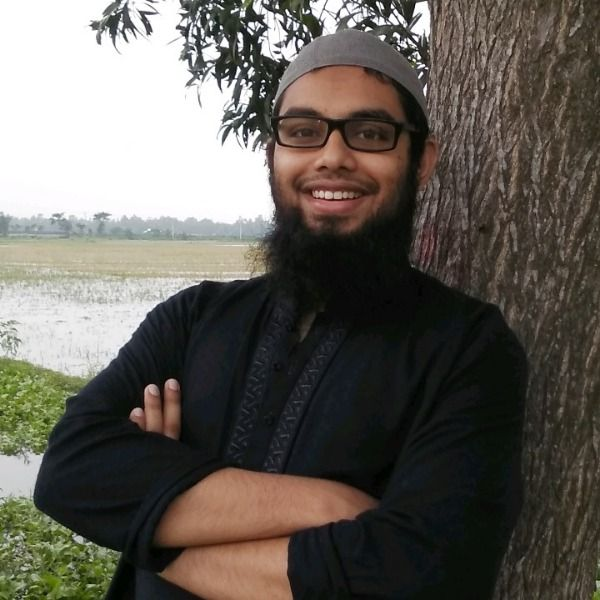}}]{Mustain Billah} is a Lecturer in the Department of Computer Science and Engineering (CSE) at Jashore University of Science and Technology (JUST), Bangladesh. Before joining at JUST, he was working as a Lecturer in the department of Information Technology (IT) at University of Information Technology and Sciences (UITS), Dhaka, Bangladesh. He also worked as Lecturer, department of Computer Science and Engineering (CSE) at Bangladesh University (BU), Dhaka.
Mustain Billah received B.Sc. (Engg.) in Information and Communication Technology (ICT) from Mawlana Bhashani Science and Technlogy University (MBSTU), Tangail and M.Sc. (Engg) in Information and Communication Technology (ICT) from the same University.
His research interest includes Data Mining, Image processing and machine Learning.
\end{IEEEbiography}

\vspace{11pt}

\begin{IEEEbiography}[{\includegraphics[width=1in,height=1.25in,clip,keepaspectratio]{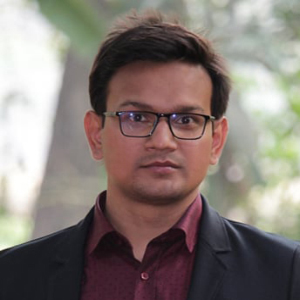}}]{Sk. Tanzir Mehedi} currently works as a Lecturer in the Department of Information Technology at the University of Information Technology and Sciences (UITS), Bangladesh. Previously, he has worked as a Data Analyst Engineer (Internship) at Fujitsu Research Institute (FRI), Japan. He has over 2 years of industrial, research, and teaching experience in universities and research laboratories. He received his B.Sc Engineering in Information and Communication Technology at the Mawlana Bhashani Science and Technology University (MBSTU), Bangladesh. Tanzir has an excellent academic background and experience in the field of Data Science and security analysis of IoT technologies with Blockchain. His research interests are in the fields of machine learning and deep learning, IoT technology, blockchain, and intrusion detection.
\end{IEEEbiography}

\vspace{11pt}

\begin{IEEEbiography}[{\includegraphics[width=1in,height=1.25in,clip,keepaspectratio]{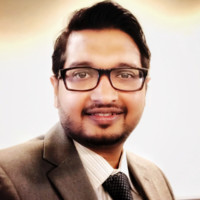}}]{Adnan Anwar}is a Cyber Security academic at Deakin University, and a member of the Centre for Cyber Security Research and Innovation (CSRI). Previously he has worked as a Data Scientist and analytics team leader at Flow Power. He has over 10 years of industrial, research, and teaching experience in universities and research laboratories including NICTA (now, Data61 of CSIRO), University of New South Wales (UNSW), La Trobe University, and Deakin University. He received his PhD and Master by Research degree from UNSW at the Australian Defence Force Academy (ADFA). He has authored over 60+ articles including journals, conference articles and book chapters in prestigious venues (H-index is 17). He has attracted research income from Government, Defence, Industries and received numerous awards at Deakin for excellence in research and teaching. Dr. Anwar’s research has greatly improved the state of the art in artificial intelligence and data-driven cybersecurity research for critical infrastructure in Australia, while his teaching is helping to develop the next generation of Australian experts (over 1200 graduates) in the area of data analytics for security and privacy.
\end{IEEEbiography}

\vspace{11pt}

\begin{IEEEbiography}[{\includegraphics[width=1in,height=1.25in,clip,keepaspectratio]{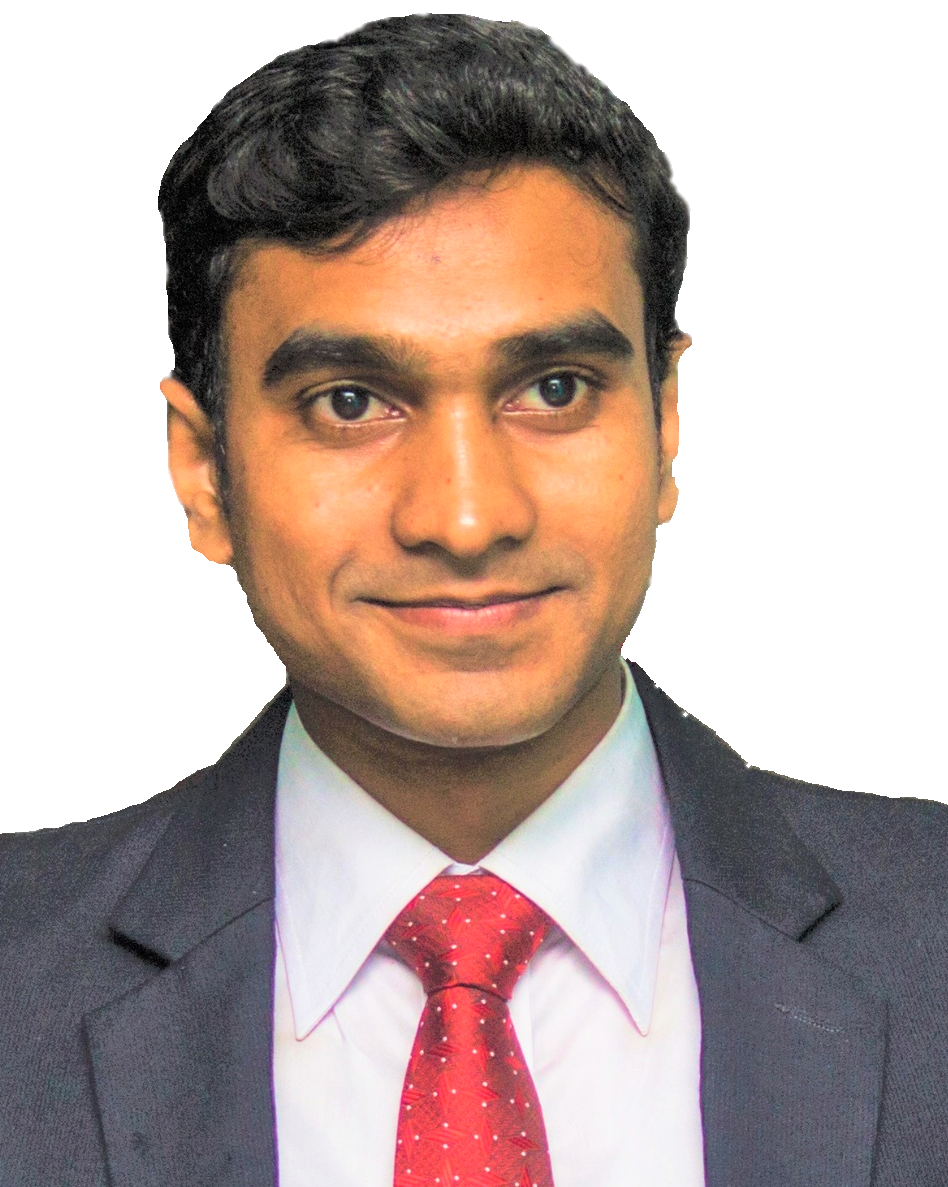}}]{Ziaur Rahman} is a Research Assistant of Deakin University, Geelong, VIC, Australia. He is PhD candidate in Cyber security of RMIT University. He served Mawlana Bhashani Science \& Technology University, Bangladesh as an Associate Professor in ICT. He casually served RMIT, Monash, Deakin and Charles Sturt University, Australia. Three (03) articles he coauthored were nominated and received the best paper awards. He is affiliated with the IEEE, ACM, Australian Computer Society. His research interests include blockchain technology, security of the internet of things (IoT), machine learning.
\end{IEEEbiography}

\vspace{11pt}

\begin{IEEEbiography}[{\includegraphics[width=1in,height=1.25in,clip,keepaspectratio]{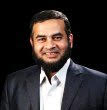}}]{Rafiqul Islam} is working as an Associate Professor at the School of Computing, Mathematics and Engineering, Charles Sturt University, Australia. Dr Islam’s main research background in cybersecurity focuses on malware analysis and classification, security in the cloud, privacy in social media, and the dark web. Dr. Islam has a strong research background in Cybersecurity with a specific focus on malware analysis and classification, Authentication, security in the cloud, privacy in social media and Internet of Things (IoT). He is leading the Cybersecurity research team and has developed a strong background in leadership, sustainability, collaborative research in the area. He has a strong publication record and has published more than 160 peer-reviewed research papers. His contribution is recognized both nationally and internationally through achieving various rewards such as professional excellence reward, research excellence award, leadership award. Dr. Islam has recognized at the national forefront of his research field ‘cybersecurity’, which is now one of the national/International research priority (financial, political and social aspects). His is Co-Investigator on a successful Cybersecurity CRC  (68 M) to which he is contributing to the projects related to Resilient Networks, Security and configuration management of IoT system, Platform \& Architecture of cybersecurity as a service and malware detection and removal.  
\end{IEEEbiography}

\EOD

\end{document}